\input harvmac

\def \const {{\rm const}}

\def \ov {\over}

\def \k {\kappa}

\def \L {{\cal L}}

\def \J {{\cal J}}

\def \pa { \partial}
\def \a {\alpha}
\def \E {{\cal E}}
\def \b {\beta}
\def \g {\gamma}

\def \l {\lambda}

\def \nn {\nabla}
\def \s {\sigma}

\def \r {\rho}

\def \ta {\tau}

\def \vp {\varphi}

\def \frac#1#2{{ #1 \over #2}}
\def \lr { \lref}

\def \aa {{\a'}}
\def \lr{\lref}

\def \rf {\refs}

\def \adss {$AdS_5 \times S^5\ $}

\def \ta { \tau}
\def \s { \sigma }

\def \vp {\varphi}

 \def \a { \alpha}
\def \r {\rho}

\def \del{\partial}

\def \ha { { 1 \over 2}}

\def \g {\gamma}

\def \k {\kappa}
\def \l {\lambda}
\def \L {{\cal L}}

\def \b{\beta}

\def \ha {{1 \over 2}}

\def \ov {\over}

\def \ww { {\rm w} }
\def \www { {\rm w}_{21} }

\def \sql {{\sqrt{\l}}\ }

\def \ta {\tau}

\def \E {{\cal E}}

\def \ta {\tau}

\def \nn {{\rm n}}

\def \kk {{\rm k}}
\def \po {{\psi_0}}

\lref\afrt{
G.~Arutyunov, S.~Frolov, J.G.~Russo and A.~A.~Tseytlin,
work in progress.
}

\lref\mets{
R.~R.~Metsaev,
``Type IIB Green-Schwarz superstring in plane wave Ramond-Ramond
background,''
Nucl.\ Phys.\ B {\bf 625}, 70 (2002)
[hep-th/0112044].
R.~R.~Metsaev and A.~A.~Tseytlin,
``Exactly solvable model of superstring in plane wave
 Ramond-Ramond
background,''
Phys.\ Rev.\ D {\bf 65}, 126004 (2002)
[hep-th/0202109].
}

\lref\beiks{
N.~Beisert, C.~Kristjansen and M.~Staudacher,
``The dilatation operator of N = 4 super Yang-Mills theory,''
hep-th/0303060.
}

\lref\beikps{
N.~Beisert, C.~Kristjansen, J.~Plefka and M.~Staudacher,
``BMN gauge theory as a quantum mechanical system,''
Phys.\ Lett.\ B {\bf 558}, 229 (2003)
[hep-th/0212269].
}

\lr \papa {
M.~Blau, J.~Figueroa-O'Farrill, C.~Hull and G.~Papadopoulos,
``A new maximally supersymmetric background of IIB superstring
 theory,''
JHEP {\bf 0201}, 047 (2002)
[hep-th/0110242].
}

\lr \bmn { D.~Berenstein, J.~Maldacena and H.~Nastase,
``Strings in flat space and pp waves from N = 4 super Yang
 Mills,''
JHEP {\bf 0204}, 013 (2002)
[hep-th/0202021].
}
\lr \gkp {
S.~S.~Gubser, I.~R.~Klebanov and A.~M.~Polyakov,
``A semi-classical limit of the gauge/string correspondence,''
Nucl.\ Phys.\ B {\bf 636}, 99 (2002)
[hep-th/0204051].
}
\lr \pol {A.~M.~Polyakov,
``Gauge fields and space-time,''
Int.\ J.\ Mod.\ Phys.\ A {\bf 17S1}, 119 (2002)
[hep-th/0110196].
}

\lr \tse { A.~A.~Tseytlin,
``Semiclassical quantization of superstrings: $AdS_5 \times S^5$
 and
 beyond,''
Int.\ J.\ Mod.\ Phys.\ A {\bf 18}, 981 (2003)
[hep-th/0209116].
``On semiclassical approximation and spinning string vertex
operators in $AdS_5 \times  S^5$,''
hep-th/0304139.
}

\lr \fts{ S.~Frolov and A.~A.~Tseytlin,
``Semiclassical quantization of rotating superstring in \adss,''
JHEP {\bf 0206}, 007 (2002)
[hep-th/0204226].
}
\lr\rus{ J.~G.~Russo,
``Anomalous dimensions in gauge theories from rotating strings in
\adss,''
JHEP {\bf 0206}, 038 (2002)
[hep-th/0205244].
}

\lr \bis{N.~Beisert, C.~Kristjansen, J.~Plefka, G.~W.~Semenoff
 and
M.~Staudacher,
``BMN correlators and operator mixing in N = 4 super Yang-Mills
theory,''
Nucl.\ Phys.\ B {\bf 650}, 125 (2003)
[hep-th/0208178].
}

\lr \mzn{N.~Beisert, J.~Minahan,
M.~Staudacher and K.~Zarembo, ``Stringing Spins and Spinning
Strings,'' [hep-th/0306139].}

\lr \mz{
J.~A.~Minahan and K.~Zarembo,
``The Bethe-ansatz for N = 4 super Yang-Mills,''
JHEP {\bf 0303}, 013 (2003)
[hep-th/0212208].
}
\lr \ft{S.~Frolov and A.~A.~Tseytlin,
``Multi-spin string solutions in $AdS_5 \times S^5$,''
hep-th/0304255.
}
\lr \ftn {S.~Frolov and A.~A.~Tseytlin,
``Quantizing three-spin string solution in $AdS_5 \times S^5$,
 hep-th/0306130.
}
\lr \gro{C.~Kristjansen, J.~Plefka, G.~W.~Semenoff and
 M.~Staudacher,
``A new double-scaling limit of N = 4 super Yang-Mills theory and
 PP-wave  strings,''
Nucl.\ Phys.\ B {\bf 643}, 3 (2002)
[hep-th/0205033].
D.~J.~Gross, A.~Mikhailov and R.~Roiban,
``Operators with large R charge in N = 4 Yang-Mills theory,''
Annals Phys.\  {\bf 301}, 31 (2002)
[hep-th/0205066].
N.~R.~Constable, D.~Z.~Freedman, M.~Headrick, S.~Minwalla,
 L.~Motl, A.~Postnikov and W.~Skiba,
``PP-wave string interactions from perturbative Yang-Mills
 theory,''
JHEP {\bf 0207}, 017 (2002)
[hep-th/0205089].
}
\lr \pol {A.~M.~Polyakov,
``Gauge fields and space-time,''
Int.\ J.\ Mod.\ Phys.\ A {\bf 17S1}, 119 (2002)
[hep-th/0110196].
}
\lr \min{J.~A.~Minahan,
``Circular semiclassical string solutions on \adss,''
Nucl.\ Phys.\ B {\bf 648}, 203 (2003)
[hep-th/0209047].
}

\Title{\vbox
{\baselineskip 10pt
{\hbox{
}
}}}
{\vbox{\vskip -30 true pt
\medskip
\centerline {
 Rotating string  solutions: }
\medskip
\centerline {
AdS/CFT duality in non-supersymmetric sectors
 }
\medskip
\vskip4pt }}
\vskip -20 true pt
\centerline{S. Frolov$^{a,}$\footnote{$^*$} {Also at Steklov
Mathematical Institute, Moscow.}
 and
A.A.~Tseytlin$^{a,b,}$\footnote{$^{**}$}
{Also at
Lebedev Physics Institute, Moscow.}
}
\smallskip\smallskip
\centerline{ $^a$ \it  Department of Physics,
 The Ohio State University,
 Columbus, OH 43210, USA}
\centerline{ $^b$ \it  Blackett Laboratory,
 Imperial College,
 London,  SW7 2BZ, U.K.}

\bigskip\bigskip
\centerline {\bf Abstract}
\baselineskip12pt
\noindent
\medskip
We find the minimal energy solution describing a folded closed
string located at the center of $AdS_5$ and rotating
simultaneously in two planes in $S^5$ with two arbitrary $SO(6)$
angular momenta $J_1$ and $J_2$. In the case  when $J_1=J_2=J'$
we observe the precise agreement between the leading coefficient
in the large $J'$ expansion of the energy of this solution and
 the minimal eigen-value of the 1-loop
anomalous dimension matrix of the corresponding  $[J',0,J']$
SYM operators  obtained by
Beisert, Minahan, Staudacher and Zarembo in hep-th/0306139.
We find also perfect agreement between string and SYM results
in cases of states with unequal spins
dual to $[J_2,J_1-J_2,J_2]$ operators. This represents a
remarkable quantitative test of the AdS/CFT duality in a
non-supersymmetric (and non-near-BPS) sector.

\bigskip
\Date{06/03}
\noblackbox
\baselineskip 16pt plus 2pt minus 2pt

\newsec{Introduction}
Precise  quantitative  checks of the AdS/CFT correspondence in
non-trivial, i.e.  non-supersymmetric,  cases
were  largely  non-existent until very recently.
A remarkable progress was initiated in \bmn,
which suggested the possibility of direct comparison
of energies of $AdS_5 \times S^5$ string
expanded near a particular BPS state
(represented by a point-like string running along a
geodesic in $S^5$)  with  perturbative dimensions
of the corresponding SYM operators.

Since  the comparison  in the BMN \bmn\  case is done for
near-BPS states, one may still  wonder if  exact numerical tests
of the AdS/CFT duality can be carried out also in a  ``far from
 BPS''
sectors, i.e. when one expands  near a non-supersymmetric  string
solution. Remarkably, the two recent developments on
the  SYM  \rf{\bis,\beikps,\mz,\mzn} and the string \rf{\ft,\ftn}
sides  have converged  to provide such non-trivial
tests.

The  adequate interpretation of the proposal of \bmn\
as a special case of a  semiclassical expansion of \adss  string
theory selecting  a particular sector of states with large
charges  was suggested  in \rf{\gkp} and developed in
 \rf{\fts,\tse}.
 Given a classical
string solution carrying some global charges (e.g., $SO(6)$
angular  momenta)\foot{More generally, semiclassical
considerations and matching onto SYM operators may apply to
a broader sector of states with large oscillation numbers
\rf{\pol,\gkp,\min}.}
and  parametrized by some constants $\ww_i$,
its classical  $AdS_5$ energy  and angular momenta $J_s$
can be written as
$E= \sql \E(\ww_i)$,  \ $J_s = \sql \J_s (\ww_i)$, where
$\sql$ is the effective string tension (or `t Hooft coupling
of the dual gauge theory). In the simplest and most interesting
cases,  some  components of $J_s$ may be equal, so let
us assume for simplicity that there is just one relevant
parameter $\ww$ and one value of $J$  and   $J= \sql \ww$.
Then the energy can be written as $E=E( J, \sql)$. The key
assumptions  that eventually allow one  to compare to  SYM theory
are:
 (i)  $ \ww \ll 1$, i.e. $g_{\rm eff}\equiv
 {1 \ov \ww^2} = { \l \ov J^2} \ll 1$,\
and (ii)  $J \gg 1$.

In the semiclassical  expansion in string theory  $ \sql \gg 1$,
while in perturbative SYM theory $\l \ll 1$,  but ${ \l \ov J^2}
\ll 1$ should be satisfied on both sides of the duality to allow
 for the
comparison. In addition, the value of $J$ should be numerically
large in  both theories. The assumption (i)  allows one to write
 the
classical  energy  as
$E_0 = k_0  J (1 + k_1 { \l \ov J^2} + ...) $,
where $k_0=1$ or 2 in the cases of interest \ft\ and dots stand
for higher powers  of ${ \l \ov J^2}$.
Other coefficients ($k_1,...$) may, in general,
be functions of ratios of unequal angular momenta (cf. \ftn).
This looks like an
expansion in positive integer powers of $\l$ and would suggest a
possibility of comparison to perturbative anomalous dimensions of
the corresponding  operators on the SYM side, if not for the fact
that  quantum  string sigma model corrections should a priori
give non-trivial contributions to $E$. It is here where the
second assumption  of $J \gg 1$ comes  into play:  it turns out
that (due to underlying supersymmetry of the \adss  superstring
 theory)
quantum sigma model corrections  to the above classical
expression for the energy are suppressed by powers of $1 \ov J$.

In more detail, on general grounds, the $\ell$-loop
sigma model correction to  the energy  should have the form
$E_\ell= { 1 \ov (\sql)^{\ell -1} } \E_\ell(\ww)$.
Assuming that $ \E_\ell = {c_\ell \ov \ww^{\l+1}} + ...$
for $\ww \gg 1$  we would get
$E_\ell= { 1 \ov J^{\ell -1} } (c_\ell { \l \ov J^2 } + ...) .$
In particular,  $E_1 = c_1 { \l \ov J^2 } + ...$ .
This is indeed what happens  in the non-trivial three-spin
example \ft\ as was shown  explicitly
at the 1-loop  order in   \ftn.
Then $E= E_0 + E_1 + ...
=  k_0  J [1 + k_1 { \l \ov J^2}( 1  + {c_1' \ov J}  + ...)
      + ... ] $,  $c_1'= {c_1 \ov k_0 k_1}$.
Assuming  that string sigma model corrections to the classical
 energy
are indeed suppressed in the large $J$ limit, one could then
expect to match the terms in the expansion of the classical
energy $E_0 = k_0 (  J  + k_1 { \l \ov J} +   k_2 { \l \ov J^3} +
...)  $ to the canonical dimension,
one-loop anomalous dimension, two-loop anomalous dimension,
etc., of the corresponding  SYM  operators
with $J \gg 1$.

That  sounds very much like the BMN story (cf. \fts\ and \gro),
where similar discussion applied to {\it excited} states near a
 BPS
state. Here instead  we are considering the {\it classical}
 energy
of a non-BPS  ``ground state'' which is ``far away'' from BPS
ground state of BMN. Matching to SYM anomalous dimensions
would be  already a non-trivial test  of the
AdS/CFT duality.  One  could   then try  to match  also
small excitations
near a given non-BPS ground state.

Examples of simple rotating   multiple-spin string solutions
for which the above considerations should apply were recently
constructed in \rf{\ft,\ftn}. These circular string solutions
have two equal angular momenta $J_1=J_2=J'$ in the two orthogonal
planes, and the center of mass of the string is rotating  in the
third plane with angular momentum $J_3=J$.  The solution with
$J_3=0$ turns  out to be unstable (for a similar reason why a
closed string wound around large circle of a sphere is unstable).
Adding extra angular momentum stabilizes the solution under small
perturbations provided $J \geq { 2 \ov 3} J'$ \ftn. \foot{A
possible interpretation of why stabilization happens
is that adding extra orbital motion
provides a centrifugal force that compensates for the contraction
caused by string tension.}

Similar  semiclassical string states with $SO(6)$ spins
$(J_1,J_2,J_3)$ should correspond to particular SYM scalar
operators tr$(X^{J_1} Y^{J_2} Z^{J_3})+ ...$  ($X,Y,Z$ are three
complex combinations of 6 real adjoint scalars) in irreducible
$SU(4)$ representations  with Dynkin labels $[J_2-J_3, J_1-J_2,
J_2 + J_3 ]$ (assuming for definiteness that $J_1 \geq  J_2 \geq
J_3$). To compare the string expressions
in \ftn\  with SYM theory  one thus needs to know
the eigen-values of perturbative  anomalous dimension
matrix  of such  operators for large values of $J_i$.

Progress in this direction was achieved very recently
in \mzn\ in the special case of $J_3$=0: the eigen-values of the
planar 1-loop anomalous dimension matrix for the
scalar SYM operators
tr$(X^{J_1} Y^{J_2})+ ...$  were computed
explicitly using integrable spin chain and dilatation
operator technique,   building on the previous work of
 \rf{\bis,\mz}.
In the special case of $J_1=J_2=J'$, i.e.
$[J', 0, J']$  operators, it was found \mzn\ that
there exists a 1-loop anomalous dimension eigen-value
that matches precisely onto the string theory prediction
for the  classical energy of a circular
 two-spin solution \ft. Moreover, the (real) energies
of fluctuation  modes near this  solution \ft\ also have  precise
counterparts among gauge theory eigen-values \mzn.

A puzzling feature of this agreement,
however, is that the two-spin circular string solution of \ft\
is unstable, i.e. there is a tachyonic fluctuation mode (not seen
on gauge theory side \mzn).  One may wonder why the comparison
should work at all   for an unstable string solution.
A possible explanation  is that the  matching  actually works for
a more general stable solution with $J_3 > { 2 \ov 3} J'$,  and
for some reason the resulting  expressions (for the energies of
the ground state as well as fluctuation  modes) on both sides of
the duality admit analytic continuation to the region  $J_3=0$.
The analytic continuation  should not  work for the would-be
tachyonic mode (whose energy and thus anomalous  dimension would
become  imaginary) and that is  why it was not seen in  \mzn.

Another surprising result  of \mzn\ is the existence of
$[J',0,J']$ eigenvalue  lower that the one corresponding to the
circular solution of \ft. Believing in AdS/CFT, this  suggests
the existence
of a  different two-spin string solution in the sector of states
 with
$J_1=J_2$
  that has energy {\it
lower} than the circular solution.

\bigskip

Our aim here  is to show   that such solution does exist and
describes {\it folded} closed string  rotating simultaneously in
the two orthogonal  planes with its center of mass  positioned
at a fixed point of $S^5$ (which is a  generalisation of the
single-spin solution in  \rf{\gkp}). We find that the
first subleading term in the  energy of this solution
$E=2J'(1+k_1{\l\ov J'^2}+ ... )$ has indeed exactly the same
coefficient as the corresponding minimal 1-loop anomalous
dimension on the SYM side! This provides a remarkable test of the
above AdS/CFT duality  considerations.

The  general  folded string solution we shall consider will
have two arbitrary angular momenta $J_1,J_2$.
We shall find its energy by expanding in $ {\l \ov (J_1+J_2)^2
 }\ll 1 $ and $ {J_2 \ov J_1 + J_2} \ll 1$, and observe again
that the first subleading term in $E$ matches precisely with the
corresponding  1-loop anomalous dimension eigenvalue for the
operator $[J_2, J_1-J_2, J_2 ]$ obtained in  \mzn.

It should be noted that,  as in many similar duality contexts,
easy-to-find  classical string energy expressions
provide predictions  for all-loop anomalous dimension expressions
on the SYM side (which are rather complicated  to  find  already
at the 1-loop order).   Such precise matching between the string
 and
SYM results  as found in \mzn\ and here  which works
for different string states suggests a more direct
equivalence  between an effective anomalous dimension
Hamiltonian (or a  kind of dilatation operator)
on the SYM side and the classical string action
(both expanded in  ${ \l \ov J^2}\ll 1$ and for $J \ll 1$).

While the  two-spin folded string solution we present  below
is a direct generalization of the single-spin solution  in
\rf{\gkp,\rus}, it  should be noted that the two-spin case
turns out to be  better suited for comparison  with the SYM
 theory
than the single-spin case -- in the latter the classical energy
has the following expansion \gkp:
$E= J + {2 \sql \ov \pi} + O({1 \ov J})$. The leading correction
 here
is much larger than its counterpart $ { \l \ov J}$ in the
two-spin case (assuming $ { \l \ov J^2}\ll 1 $), and thus its
direct comparison to SYM theory is problematic (it should also
receive corrections in string sigma model loop  expansion).
The point is that in the single-spin sector the minimal-energy
state is represented by the BMN  point-like string  dual to
the   BPS gauge-theory operator tr$Z^J$,  while other states
having higher energy should be represented by operators with
extra  zero R-charge insertions  like tr$( Z^J X \bar X ...
)$ or tr$[ Z^J  (F_{mn})^2 ... ]$. At the same time,
the folded string solution appears to represent the minimal
energy configuration in the two-spin sector  (where there is
apparently no BPS state) and that  is why it can be matched onto
the minimal dimension  operators like
tr$(X^{J_1} Y^{J_2})+...$.

\newsec{
Two-spin string solution in $AdS_5\times
S^5$}
Let us start with specifying our notation.
The bosonic part of the \adss string action is
\eqn\gsss{
I= - { \sql  \ov 4\pi }
\int d^2 \xi  \ \big[ G^{(AdS_5)}_{mn}(x)
\del_a x^m \del^a  x^n\ + \    G^{(S^5)}_{pq}(y)  \del_a y^p
\del^a y^q \big] \ , \ \ \ \ \ \ \ \ \sql \equiv  { R^2 \ov
\aa} \ .}
We shall use the following explicit parametrization of
the unit-radius  metric on $S^5$:
\eqn\Sd{
(ds^2)_{S^5}
= d\g^2 + \cos^2\g\ d\vp_3^2 +\sin^2\g\ (d\psi^2 +
\cos^2\psi\ d\vp_1^2+ \sin^2\psi\ d\vp_2^2)\ .}
It will be convenient to choose the range of $\psi$ to be the
interval $[-\pi, \pi ]$.
This metric has three translational
isometries in $\vp_i$, so that
 in addition to the three $AdS_5$
integrals of motion,
 a general solution should also
have  the following  three
 integrals of motion depending on the
$S^5$ part of the action:
\eqn\jiii{ J_3= {\sql  } \int^{2\pi}_0
{d \s\ov  2 \pi} \ \cos^2 \g\ \pa_0\vp_3 \equiv \sql \J_3
\ ,   }
\eqn\ji{
J_1= {\sql   } \int^{2\pi}_0 {d \s\ov  2 \pi}
 \ \sin^2 \g \ \cos^2 \psi \ \pa_0 \vp_1 \equiv \sql \J_1
\ , }
\eqn\jii{
J_2= {\sql   } \int^{2\pi}_0 {d \s\ov  2 \pi}
 \ \sin^2 \g \ \sin^2 \psi \ \pa_0 \vp_2 \equiv \sql \J_2
   \ .   }
Let us look for a classical solution describing a closed folded
 string
located at the center $\r = 0$
of $AdS_5$  and at fixed  value of the $S^5$ angle
 $\g={\pi \ov 2} $, rotating  within $S^3$ part of $S^5$  with
arbitrary frequencies $\ww_1$ and $\ww_2$. A natural ansatz for
such a solution is
\eqn\ann{  t= \k \ta \ , \ \ \ \  \r =0 \ , \
\ \ \g={\pi \ov2}  \ , \ \ \ \
 \vp_3= 0 \ , \ \ \
 \ \ \vp_1 = \ww_1 \tau \ ,  \ \ \ \ \  \vp_2 = \ww_2
\tau \ ,  \ \ \ \ \ \psi= \psi(\s)  \ , }
where $  \k,\ \ww_1,\ \ww_2 =\const$.
Such solution will have $J_3=0, \ J_1\not=0, \ J_2\not=0$.
The energy  is
\eqn\enn{ E= \sql \k = E(J_1,J_2, \l)  \ . }
The special  case of $\ww_1=0$  ($J_1=0$)
will give  the single-spin
folded string solution considered in \gkp.
One of our
aims will be to find a minimal energy solution in the sector
of semiclassical string configurations with
$J_1=J_2$.

We will assume without loss of generality that $\ww_2 \ge
\ww_1$ and introduce
$$
\www^2 \equiv  \ww_2^2 - \ww_1^2 \geq 0 \ .
$$
The string equations of motion for the angles
in the conformal gauge  then lead to the
following equation for $\psi$
\eqn\eqpsi{
\psi'' + \ha\www^2 \sin 2\psi =0\ .}
Integrating this once, we get
\eqn\eqpsii{
\psi'^2 =\www^2(\sin^2\po - \sin^2\psi ) + \kk^2  \ ,
}
where the integration constant
$\po$ will be  determining  the length of a  folded string
as in \gkp.
We have introduced an (integer) parameter $\kk$  to
include also  the case when $\ww_1=\ww_2$.
The conformal gauge constraints lead to the following relation
between $\k$, $x_0$ and $\ww_i$
\eqn\energy{
\k^2 = \psi'^2 + \ww^2_1 \cos^2\psi +
\ww^2_2\sin^2\psi = \ww_2^2 \sin^2 \po +
\ww_1^2 \cos^2 \po + \kk^2 \ . }
There are two distinct solutions,  the circular (C) one
 \ft\  and the  new folded (F)  one (where $\kk=0$):\foot{
As in the single-spin case \gkp,
the solution with both $\www$ and $\kk$ being non-zero
(i.e. a
generalization of the  circular string solution)
 is  expected to  be unstable.}
\eqn\ass{
C:  \ \ \ \
\ww_1=\ww_2=\ww\ , \ \ \ \ \ \ \ \psi =\kk \s \ , \  \ \ \ \ \
\k^2 = \ww^2 + \kk^2 \ , \ \ \ \  \kk=1,2,... \ , }
\eqn\asds{
F:  \ \ \ \
\ww_1\not =\ww_2\  , \ \ \ \ \ \ \ \psi =\psi( \s) \ ,\ \
-\psi_0\le \psi( \s)\le\psi_0\ ,   \  \ \ \ \ \ \k^2 = \ww_2^2
\sin^2 \po + \ww_1^2 \cos^2 \po   \ . }
For the single-fold solution the
 periodicity in $\s$ implies the following
condition on the parameters:
\eqn\period{
2 \pi = \int^{2\pi}_0 d \s
= 4 \int^{\psi_0}_0   { d \psi \ov \www\sqrt{  \sin^2 \po  -
 \sin^2\psi  }   } \ .}
The angular momenta are
\eqn\jin{
\J_1= \ww_1\int^{2\pi}_0 {d \s\ov  2 \pi}
 \ \cos^2 \psi = {2\ww_1\ov \pi\www}\int^{\psi_0}_0
 {\cos^2 \psi\, d \psi \ov
\sqrt{  \sin^2\psi_0 - \sin^2\psi  }   }\ , }
\eqn\jiin{
\J_2= \ww_2 \int^{2\pi}_0 {d \s\ov  2 \pi}
 \ \sin^2 \psi  = {2\ww_2\ov \pi\www}\int^{\psi_0}_0
 {\sin^2 \psi\, d \psi \ov
\sqrt{  \sin^2\psi_0- \sin^2\psi  }   }\ .   }
Note the following relation
\eqn\reljw{
{\J_1\ov\ww_1} + {\J_2\ov\ww_2} = 1\ .}
Computing the integrals, we get the following expressions in
 terms of the hypergeometric functions
\eqn\periodi{
\www = {}_2F_1\left({1\ov 2},{1\ov 2};1;x_0\right)\ , \ \ \ \ \ \
  \ \
x_0 \equiv  \sin^2 \po \ , }
\eqn\jini{
\J_1={\ww_1\ov\www}\, {}_2F_1\left(-{1\ov 2},{1\ov
2};1;x_0\right)\ , \ \ \ \ \
\J_2={\ww_2x_0\ov 2\www}\, {}_2F_1\left({1\ov 2},{3\ov
2};2;x_0\right)\ .}
These relations can be used to find the dependence of $\ww_i$
and $\psi_0$ on the angular momenta $J_1,J_2$, and then, using
 \enn,\energy\
one can find the dependence of the energy on
the angular momenta.

\newsec{ Sector of equal spins}
Let us now consider the  case
$$                  J_1 = J_2 = J'  \ .   $$
In the  circular solution case we get (here $\kk=1$) \ft\
\eqn\irc{
E=  2 J' \sqrt{ 1 + { \l \ov (2J')^2}}\ . }
In the case of the   folded string solution, which  happens
to have lower (minimal) energy in this  sector,
the expression for the energy $E=E(J',\l) $ can be
found in expansion in powers of
 $${ 1 \ov \J^2 } = { \l  \ov J'^2  } \ll 1   \ . $$
 Solving \reljw, \periodi\ and \jini, we find
\eqn\xo{
x_0 = 0.826115 - \frac{0.0784101}{\J^2}+ \frac{0.00697882}{\J^4}
+ ...\ ,}
\eqn\wwi{
\ww_1 = 2\J - \frac{0.272922}{\J}+ \frac{0.0858257}{\J^3}
+ ... \ , \ \ \ \ \ \ \
\ww_2 = 2\J +\ \frac{0.272922}{\J} \ - \ \frac{0.0113391}{\J^3}
+ ... \ ,}
\eqn\gyi{
E  = 2 J'+ k_1 \frac{ \l}{4J'} \  -  \
k_2 \frac{ \l^2}{ 64 J'^3} + ... \ , }
\eqn\coff{
k_1 = 0.712032 \ ,\ \ \ \ \ \ \ \ k_2 = 1.69878 \ . }
For comparison, for the circular string solution \irc\ we find
\eqn\rgyi{
E = 2 J'+ \frac{ \l}{4J'} \  -  \ \frac{ \l^2}{64  J'^3} + ... \
 .}
While  the numerical coefficients in \gyi\ are
of order 1  as in \rgyi,  the energy of the folded string
 solution \gyi\
is clearly {\it lower}  than the energy of the circular solution
for a given value of the angular momentum $J' \gg \sql $.
The expression \gyi\ applies to the single-fold solution;
the result for the $\nn$-fold solution  turns out to be
the same but with
$\l $  replaced  by $ \l \nn^2$, i.e.
\eqn\gyil{
E  = 2 J'+ k_1 \frac{ \l \nn^2}{4J'} \  -  \
k_2 \frac{ \l^2\nn^4 }{ 64 J'^3} + ... \ ,  \ \ \ \ \ \ \
\nn= 1,2, ... \ . }
Similarly, one can generalize \rgyi\ by introducing the
``winding number'' $\kk$ in \ass\  by $\l \to \l \kk^2$ (see \ftn).
Thus the double-fold solution has higher energy than
the singly-wound  circular string, but still lower energy than
the doubly-wound string, etc.

As was found in \mzn, the subleading term in \rgyi\
can be matched precisely onto  an eigen-state of
 1-loop anomalous dimension operator for
the scalar $[J',0,J']$ operators  on the SYM side.
Ref.  \mzn\  discovered also a
{\it lower}  anomalous dimension eigen-value for which there was
 no apparent  string theory counterpart.
It was suggested  in \mzn\ that there should be a
 dual semiclassical solution which
should then involve elliptic integrals in order to match the Bethe
ansatz prediction for the anomalous dimension.

This is indeed what we find here:
the
minimal-eigenvalue  gauge-theory operator should be  dual to the
minimal-energy two-spin folded string solution.
Comparing the second term in $E$  in \gyi\ with the
result for the minimal value of the  1-loop anomalous dimension
of the $[J',0,J']$ scalar operator found  in
 \mzn\  we observe the  exact  agreement  of the
numerical coefficient $k_1$ with the coefficient
in eq. (1.7) in \mzn!\foot{Both numerical coefficients come
 from  values of certain hypergeometric functions
that can  probably be  matched directly.}

The third term in \gyi\ represents
 a prediction for the {\it two-loop}
perturbative correction to the dimension of the
SYM  $[J',0,J']$ scalar operator. It is plausible that this
prediction can be verified  by using the results of  \beiks.

\newsec{Case of unequal spins}
If the spins  $J_1,J_2$
are arbitrary,  there are two interesting cases to consider.
The first case corresponds to
$$ J_2 \ll L\ ,\ \ \ \ \ \
L \equiv  J_1+J_2 \ ,\ \ \ \ \ \ {\rm  i.e.} \ \ \ \ \
\a  \equiv
{J_2\ov L} \ll 1  \ . $$
Such string states should be
dual to  SYM  operators close to the BPS
operator in the $[0,J_1,0]$ representation
($J_1 \gg J_2$).

The second
 case is the one of almost equal spins, i.e.
$$  S \ll  L  \ ,\ \ \ \ \ S\equiv J_1 - J_2 \ .
$$
Such string states should be
dual to  SYM  operators close to the operator in
the $[J',0,J']$ irrep.

\subsec{ $\a\ll 1$}

 Solving \reljw, \periodi\ and \jini, and introducing $\L =
{L\ov \sqrt{\l}}$, we find the following expansions up to the
order $\a^8$
$$x_0 =  2\,\a\big( 1 - \frac{\a}{4} -
\frac{{\a}^2}{8} - \frac{11\,{\a}^3}{128} -
\frac{17\,{\a}^4}{256} - \frac{55\,{\a}^5}{1024} -
\frac{179\,{\a}^6}{4096} - \frac{9061\,{\a}^7}{262144}\big)
$$ \eqn\xoaa{
-{\a\ov \L^2}\big( 1 - \frac{\a}{2} -
\frac{{\a}^2}{4} - \frac{7\,{\a}^3}{32} -
\frac{119\,{\a}^4}{512} -
\frac{269\,{\a}^5}{1024} -
\frac{307\,{\a}^6}{1024} -
\frac{43165\,{\a}^7}{32768}\big)  }
$$
+{3\a\ov 4\L^4}\big( 1 - \a - \frac{3\,{\a}^2}{8} -
\frac{5\,{\a}^3}{16} -   \frac{47\,{\a}^4}{128} -
\frac{63\,{\a}^5}{128} - \frac{2799\,{\a}^6}{4096} +
\frac{39151\,{\a}^7}{32768}\big) + ...\ ,
$$
\eqn\wwiaa{ \ww_1 = \L - {\a\ov 2\,\L}\big( 1 + \a +
\frac{9\,{\a}^2}{8} + \frac{21\,{\a}^3}{16} +
\frac{795\,{\a}^4}{512} + \frac{945\,{\a}^5}{512} +
\frac{2247\,{\a}^6}{1024} +   \frac{26589\,{\a}^7}{16384} \big)
}
$$
+{3\a\ov 8\,\L^3}\big( 1 + \frac{4\,\a}{3} +
\frac{25\,{\a}^2}{12} + \frac{13\,{\a}^3}{4} +
\frac{1267\,{\a}^4}{256} + \frac{235\,{\a}^5}{32} +
\frac{21789\,{\a}^6}{2048} + \frac{564453\,{\a}^7}{32768}
\big) + ... \ ,$$
\eqn\wwiiaa{
\ww_2 = \L+   {1 \ov 2\,\L} \big( 1 + \frac{{\a}^2}{8} +
\frac{3\,{\a}^3}{16} + \frac{123\,{\a}^4}{512} +
\frac{75\,{\a}^5}{256} + \frac{357\,{\a}^6}{1024} -
\frac{9363\,{\a}^7}{16384} - \frac{26589\,{\a}^8}{16384}\big) }
$$
-{1 \ov 8\,\L^3} \big( 1 - \a - \frac{{\a}^2}{4} +
\frac{{\a}^3}{4} +   \frac{219\,{\a}^4}{256} +
\frac{429\,{\a}^5}{256} + \frac{5721\,{\a}^6}{2048} +
\frac{479271\,{\a}^7}{32768} -
\frac{6662229\,{\a}^8}{131072}\big) + ... \ . $$
The resulting
expression for the energy is
\eqn\gyiaa{
E  = L + m_1 {\a\l \ov 2L}  - m_2 {\a\l^2 \ov 8L^3} + ... \ , }
\eqn\asssq{
m_1=  1 + \frac{\a}{2}  +
\frac{3\,{\a}^2}{8} +
\frac{21\,{\a}^3}{64} +
\frac{159\,{\a}^4}{512} +
\frac{315\,{\a}^5}{1024} +
\frac{321\,{\a}^6}{1024} + ...\ ,  }
\eqn\fguy{ m_2= 1 + 2\,\a +
\frac{9\,{\a}^2}{4} + \frac{21\,{\a}^3}{8} +
\frac{807\,{\a}^4}{256} + \frac{123\,{\a}^5}{32} +
\frac{9663\,{\a}^6}{2048} + ... \ . }
Remarkably, the  expression for the coefficient $m_1$
\asssq\ in the energy  is exactly the same as
the corresponding 1-loop anomalous dimension eigenvalue
in   eq. (2.21) in \mzn !
The $m_2$-term in \gyiaa\ is  a
prediction for the  two-loop anomalous dimension.
The first terms (one) in $m_1$ and $m_2$ are consistent with the
expansion of the pp-wave \papa\ spectrum \mets\ at small $\a$.

\subsec{ $S\ll L$}
Solving \reljw, \periodi\ and \jini, and introducing $\b = {S\ov
L}$,  we find up to the order $\b^3$
\eqn\xoa{
x_0 =0.83 - 0.57\,\beta - 0.37\,{\beta}^2 +
0.19\,{\beta}^3 \  }
$$
-\frac{0.31}{\L^2}\big( 1 - 1.85\,{\beta}^2 +
1.74\,{\beta}^3\big)
+\frac{0.11}{\L^4}\big( 1 +  3.07\,\beta -
8.55\,{\beta}^2 + 11.46\,{\beta}^3 \big) + ...  \ ,$$
\eqn\wwia{
\ww_1 =\L - \frac{0.55}{\L}\big( 1 - 2.30\,\b +
3.16\,\b^2 - 4.46\,\b^3\big) +
\frac{0.69}{\L^3}\big( 1 - 3.61\,\b + 8.67\,\b^2 -
18.46\,\b^3\big) + ...  \ , }
\eqn\wwiia{
\ww_2 =  \L + \frac{0.56}{\L}\big( 1 - 0.30\,\b +
0.56\,\b^2 - 0.76\,\b^3\big) -
\frac{0.09}{\L^3}\big( 1 - 1.61\,\b + 7.72\,\b^2 -
17.60\,\b^3\big) + ... \ ,} and the energy is
\eqn\gyia{
E  =  L+ d_1 \frac{\l }{L}  + d_2  \frac{\l^2 }{L^3} +
...
\ , }
\eqn\zsdfi{
d_1= 0.356016 - 0.545845\,\b +  0.356016\,\b^2 - 0.336751\,\b^3 +
... \ ,}
\eqn\zsdfii{
d_2 = -0.212347 + 0.582988\,\b - 0.931684\,\b^2 + 1.42221\,\b^3 +
... \ . }
The first two terms in the expression for $d_1$
coincide with the corresponding 1-loop anomalous dimension
obtained in Section 2 of \mzn,
i.e. we find again a perfect agreement.
The expression for $d_2$ represents the string theory prediction
for the 2-loop perturbative anomalous dimension.

\newsec{Concluding remarks  }
There are several possible generalizations.
We  can  try to  generalize the above  2-spin
folded string solution
by introducing   one extra  angular momentum component $J_3$
 adding  the   rotation in $\vp_3$ angle.
In contrast to the circular solution case in \rf{\fts,\ftn}
where we could keep
the angle $\g$ constant, here it should be made $\s$-dependent:
\eqn\ynn{  t= \k \ta \ , \  \
 \ \ \vp_1 =  \ww_1\tau \ , \ \ \ \
 \vp_2 =\ww_2 \tau \ ,  \ \ \ \ \
\vp_3= \ww_3 \tau \ , \ \ \ \ \ \g=\g(\s) \ , \ \
\  \  \psi= \psi(\s ) \ . }
The resulting equation for $\psi$ becomes (cf. \eqpsi)
\eqn\deqw{
 (\sin^2 \g \ \psi')'  +  \ha (\ww_2^2 - \ww_1^2)  \sin^2 \g\
\sin 2\psi =0\ , }
while the equation for $\g$ follows from the conformal gauge
 constraint
(cf. \energy)
\eqn\ugy{
  \g'^2 + \ww_3^2  \cos^2 \g
+ \sin^2 \g \ ( \psi'^2 + \ww^2_1 \cos^2\psi + \ww^2_2\sin^2\psi)
 = \k^2  \ . }
We  leave the analysis of this case
(which is related to an integrable model)
 for the future work \afrt.
One can also consider similar minimal-energy  two-spin  solutions
in $AdS_5$, generalizing single-spin solution considered in  \gkp.
The challenge then will be to find the SYM
predictions for the corresponding anomalous dimensions
to allow for a comparison.

\bigskip\bigskip
\noindent
{\bf Acknowledgements}

\noindent
We are grateful to J. Russo and K. Zarembo for
useful discussions. We also thank N. Beisert, J. Minahan, M.
Staudacher and K. Zarembo for informing us about ref. \mzn\ prior
to publication. This work  was supported by the DOE grant
DE-FG02-91ER40690. The work of A.T.  was also supported in part
by the  PPARC SPG 00613 and  INTAS  99-1590 grants and the Royal
Society  Wolfson award.
\vfill\eject
\listrefs
\end